\documentclass[aps,preprint,prb,showpacs,showkeys]{revtex4-1}%
\usepackage{amssymb}
\usepackage{amsfonts}
\usepackage{amsmath}
\usepackage{graphicx}%
\providecommand{\U}[1]{\protect\rule{.1in}{.1in}}

\begin{document}
\preprint{REV\TeX4-1 }
\title{Generalized Centripetal Force Law and Quantization of Motion Constrained on
$2D$ Surfaces}
\author{Q. H. Liu}
\email{quanhuiliu@gmail.com}
\affiliation{School for Theoretical Physics, School of Physics and Electronics, Hunan
University, Changsha 410082, China}
\date{\today}

\begin{abstract}
For a particle moves on a $2D$ surface $f(\mathbf{x})=0$ embedded in $3D$
Euclidean space, the geometric momentum and potential are simultaneously
admissible within the Dirac canonical quantization scheme for constrained
motion. In our approach, not the full scheme but the symmetries indicated by
classical brackets $[x,H]_{D}$ and $[p,H]_{D}$ in addition to the fundamental
ones $[\mathbf{x},\mathbf{x}]_{D}$, $[\mathbf{x},\mathbf{p}]_{D}$ and
$[\mathbf{p},\mathbf{p}]_{D}$ are utilized, where the subscript $D$ stands for
the Dirac bracket. The generalized centripetal force law $\mathbf{\dot{p}%
=}[\mathbf{p},H]_{D}$ for particle on the $2D$ surface play the key role, and
there is no simple relationship between the force on a point of the surface
and its curvatures of the point, in sharp contrast to the motion on a curve.

\end{abstract}

\maketitle

\textit{Introduction} For the motion constrained on a $2D$ surface
$\Sigma\subseteq E^{3}$ described by an implicit equation $f(\mathbf{x})=0$
where $\mathbf{x}$ are the usual Cartesian coordinates, there are
extrinsic-curvature dependent geometric momentum and potential,
\cite{jk,dacosta,FC,liu11-1,liu13-1}\ resulting from the so-called confining
technique. By the confining technique,\ the surface is not a mathematically
surface without thickness but a $3D$ object in $E^{3}$ with e.g. thickness of
at least one atom, thus we can first imagine that there is a confining
potential crossing the surface, e.g., a harmonic potential $\mu\omega^{2}%
b^{2}/2$ with $\mu$ denoting the mass and $b$ being the normal coordinate of
the surface, then let the confining strength $\omega$ be so larger that the
motion constrained on the surface $f(\mathbf{x})=0$ is realized. The resultant
geometric potential was recently experimentally confirmed. \cite{exp1,exp2}
However, there is a difficulty: if we start the surface equation
$f(\mathbf{x})=0$ and work within the formalism of Dirac's theory of
constrained systems,\ \cite{dirac1,dirac2} the desired form of the geometric
potential appears to be unattainable. \cite{homma,ikegami,matsutani,Kleinert}
The known resolution was to resort to another but derived form of constraint
$df(\mathbf{x})/dt=0$ with the consistent form of the classical Hamiltonian
which differs from the usual one $H_{c}=\mathbf{p}^{2}/2\mu$, even for
simplicity only the kinetic energy is considered. Homma group and Ikegami
group have independently developed a formalism that opens a wide door to
various forms of the curvature-induced potentials that contain the geometric
one as a special case. \cite{homma,ikegami} Furthermore, Matsutani examined
both the classical and quantum mechanics for motion constrained on the $2D$
surfaces, and concluded that the constraint $f(\mathbf{x})=0$ is not physical
at all. \cite{matsutani} However, during recent years, we dealt with some $2D$
surfaces, case-by-case, and found that the constraint $f(\mathbf{x})=0$ is
truly physical as well, if not more. \cite{liu11-1,liu13-3,liu14-1} In the
present work, we report a universal way to resolve the problem.

There is a working hypothesis that Hamiltonian operator for a system on $2D$
surfaces also takes the usual form $H=\mathbf{p}^{2}/2\mu$ with proper form of
the momenta $\mathbf{p=(}p_{x},p_{y},p_{z}\mathbf{)}$, and it is unfortunately
too much widely accepted. \cite{homma,ikegami,matsutani,Kleinert,weinberg} In
flat space this hypothesis works, because it is nothing but a consequence of
the full Dirac canonical quantization scheme, \cite{dirac3} which formally
states that all symmetries expressed by the Poisson brackets $[\alpha
,\beta]_{P}$ between any pair of two classical quantities $\alpha$\textbf{
}and $\beta$ persist in quantum mechanics. \cite{dirac1,dirac2,dirac3} So,
once the working hypothesis meets with difficulty or fails, we have to turn to
the fundamental principles. In order to get the quantum Hamiltonian without
invoking the the full Dirac canonical quantization scheme, we should require
that $[\mathbf{x},H_{c}]_{D}$ and $[\mathbf{p},H_{c}]_{D}$, where
$[\alpha,\beta]_{D}$ denotes that Dirac bracket, persist in quantum mechanics.
This is the minimum \textit{enlargement} of the quantization rule from the
fundamental ones among $\mathbf{x}$ and $\mathbf{p}$, i.e., $\left[
\mathbf{x,x}\right]  $, $\left[  \mathbf{x,p}\right]  $ and $\left[
\mathbf{p,p}\right]  $ to include $[\mathbf{x},H]$ and $[\mathbf{p},H]$. The
Hamiltonian operator is simultaneously determined by commutation relations and
$[\mathbf{x,}H]\equiv i\hbar O\left(  [\mathbf{x},H_{c}]_{D}\right)  $ and
$[\mathbf{p,}H]\equiv i\hbar O\left(  [\mathbf{p},H_{c}]_{D}\right)  $,
provided that there is classical Hamiltonian $H_{c}$, where $O\left(
\alpha\right)  $ is the proper form of the quantum operator representing the
classical quantity $\alpha$. This is the so-called an \textit{enlarged
canonical quantization scheme }for the constrained motion on the hypersurface.
\cite{liu11-1,liu13-2,liu15,note} This scheme leads to the quantum Hamiltonian
$H=\mathbf{p}^{2}/2\mu$ for the free motion in flat space. In addition, it
leads to the geometric momentum ${\mathbf{p}}=-i\hbar({\nabla_{\Sigma}%
}+{M{\mathbf{n/}}}2)$ \cite{liu13-2,note1} where ${\nabla_{\Sigma}}$ is the
gradient operator defined on the hypersurface, and ${{\mathbf{n}}}$ is the
unit normal vector, and the mean curvature ${M}$ is usually defined by the sum
of all principal curvatures on the hypersurface. But for a $2D$ surface, the
mean curvature is usually defined by the true average so we use ${\mathbf{p}%
}=-i\hbar({\nabla_{\Sigma}}+{M{\mathbf{n}}})$ in the rest part of the Letter.

In whole of this study, we consider the free motion only without involving the
external forces which can be simply treated if no coupling between the
external forces and the curvature of the surface. After quantization, the
quantum free motion Hamiltonian on the surface has curvature-induced geometric
momentum potential $V_{g}\equiv-\mathbf{\hbar}^{2}\left(  M^{2}-K\right)
/(2\mu)$, \cite{jk,dacosta,FC,liu11-1,liu13-1}
\begin{equation}
H_{c}=\frac{\mathbf{p}^{2}}{2\mu}\rightarrow H=-\frac{\mathbf{\hbar}^{2}}%
{2\mu}\Delta_{LB}+V_{g}=-\frac{\mathbf{\hbar}^{2}}{2\mu}\left(  \Delta
_{LB}+M^{2}-K\right)  \label{1}%
\end{equation}
where $\Delta_{LB}$ is the Laplace-Betrami operator on the surface, and $K$ is
the Gaussian curvature and it is zero for the cylinder, cone, etc..

For the particle moving on the $2D$ surface $f(\mathbf{x})=0$, the unit normal
vector is $\mathbf{n=}\nabla f/\left\vert \nabla f\right\vert $. In classical
mechanics, we have for the time derivative $\mathbf{\dot{p}\equiv}%
d\mathbf{p}/dt$ of momentum $\mathbf{p}$, \cite{ikegami,weinberg,note2}%
\begin{equation}
\mathbf{\dot{p}=-np\cdot\nabla n\cdot p/}\mu.\label{2}%
\end{equation}
It in fact expresses the \textit{generalized} \textit{centripetal force law}
(GCFL) which should reduce to the \textit{usual one} $a=v^{2}/R$ as the
particle is constrained on a curve, so this relation must then be in general
expressible as those between kinematic quantities and intrinsic/extrinsic
curvatures. With noting $\mathbf{\dot{p}=}\left[  \mathbf{p,}H_{c}\right]
_{D}$, the \textit{enlarged canonical quantization scheme} implies that the
following relation holds true during quantization,
\begin{equation}
\left[  \mathbf{p,}H\right]  =-i\hbar O\left(  \mathbf{np\cdot\nabla n\cdot
p/}\mu\right)  ,\label{3}%
\end{equation}
The key finding of this Letter is that the geometric momentum and potential
can automatically appear in this relation (\ref{3}). We will first deal with
some special $2D$ surfaces then make remarks on the general one.

\textit{Case 1: motion on cylinders} By the $2D$ cylinder we mean a ruled
surface spanned by a one-parameter family of parallel lines along $z$-axis for
convenience. So the cross section of the cylinder can be an ellipse, a
parabola, a hyperbola, and a curved or even a straight line, and their
equations can be assumed to be given by $y=u(x)$ whose curvatures take the
form $\kappa=$ $1/R=u^{\prime\prime}(x)/(1+u^{\prime}(x)^{2})^{3/2}$. It is
easily understandable that GCFL (\ref{2}) takes the following form\ for it is
nothing but another form of well-known one $a=v^{2}/R$,
\begin{equation}
\mathbf{\dot{p}}=-4\mathbf{M}H_{c}=2H_{c}\mathbf{n}/R,\text{ or, }\left[
\mathbf{p,}H_{c}\right]  _{D}=2H_{c}\mathbf{n}/R,\label{4}%
\end{equation}
where $H_{c}=\mu v^{2}/2=\left(  p_{x}^{2}+p_{y}^{2}\right)  /2\mu$ is the
classical Hamiltonian for the motion on the the cross section for the motion
along axis of the cylinder is trivial thus is neglected, $\mathbf{M=-n}%
/\left(  2R\right)  $ is the mean curvature vector, a geometric invariant, and
$\mathbf{n}$ being the normal vector. To note that, in differential geometry,
mean curvature is defined by $M=-1/\left(  2R\right)  $ whose sign depends on
the choice of normal, negative if the normal points along the convex side of
the surface. Equation\ (\ref{4}) shows that the generalized centripetral force
is proportional to the mean curvature.

In quantum mechanics, Eq. (\ref{3}) is now,
\begin{equation}
\left[  \mathbf{p,}H\right]  =i\hbar O\left(  2H_{c}\mathbf{n}/R\right)
.\label{5}%
\end{equation}
Previous studies demonstrate that no matter what form of the momentum is
taken, the Hamiltonian operator $H=\mathbf{p}^{2}/2\mu$ is not able to include
the geometric potential. \cite{homma,ikegami,matsutani} Instead, it is an easy
task to establish the following equations for unknown functions $q_{i}(x)$
($i=x,y$),%
\begin{equation}
\left[  p_{i}\mathbf{,}H\right]  =i\hbar\left(  e^{-q_{i}(x)}\frac{n_{i}}%
{R}He^{q_{i}(x)}+e^{q_{i}(x)}H\frac{n_{i}}{R}e^{-q_{i}(x)}\right)  ,\text{
}(i=x,y),\label{6}%
\end{equation}
which in classical limit reduces to the classical one (\ref{4}) because the
factors $e^{\pm q_{i}(x)}$ cancel or becomes dummy. \cite{Kleinertbook} Eqs.
(\ref{6}) have explicit forms with recalling $\kappa=1/R$,
\begin{align}
\left(  \frac{dq_{x}}{dx}\right)  ^{2}+\frac{u^{\prime}\kappa^{\prime}%
-g\kappa^{2}}{\kappa u^{\prime}}\frac{dq_{x}}{dx}+\frac{u^{\prime}%
\kappa^{\prime\prime}+2g\kappa\kappa^{\prime}-gu^{\prime2}\kappa\kappa
^{\prime}}{4\kappa u^{\prime}} &  =0,\label{7-1}\\
\left(  \frac{dq_{y}}{dx}\right)  ^{2}+\frac{g\kappa^{2}u^{\prime}%
-\kappa^{\prime}}{\kappa}\frac{dq_{y}}{dx}+\frac{\kappa^{\prime\prime
}-3gu^{\prime}\kappa\kappa^{\prime}}{4\kappa} &  =0,\label{7-2}%
\end{align}
where $g=\sqrt{1+u^{\prime}(x)^{2}}$ is the determinant of the metric tensor
or the length of a normal vector $\nabla f(\mathbf{x})$. Though for some
important cases we can obtain the closed form solutions for $q_{i}(x)$, e.g.,
$q_{i}(x)=0$ for the cross section is a circle, \cite{liu13-2} etc., they are
in general unavailable.

In fact, there are simpler equations with closed form solutions\ available for
$q_{i}(x)$ for ordinary curves $y=u(x)$. Here we report such one,
\begin{equation}
\left[  p_{i}\mathbf{,}H\right]  =i\hbar\frac{1}{3}\left(
part1+part2+part3\right)  ,\text{ }(i=x,y),\label{8}%
\end{equation}
where all $part1$, $part2$ and $part3$ in classical limit reduce to
$2H_{c}n_{i}/R$, and their explicit forms are, respectively,
\begin{align}
part1 &  =e^{-\sqrt{2}q_{i}(x)}\frac{n_{i}}{R}He^{\sqrt{2}q_{i}(x)}%
+e^{\sqrt{2}q_{i}(x)}H\frac{n_{i}}{R}e^{-\sqrt{2}q_{i}(x)},\label{9-1}\\
part2 &  =\frac{n_{i}}{R}H^{\prime}+H^{\prime}\frac{n_{i}}{R},\text{
(}H^{\prime}=\left(  e^{-q_{i}(x)}p_{x}e^{2q_{i}(x)}p_{x}e^{-q_{i}%
(x)}+e^{-q_{i}(x)}p_{y}e^{2q_{i}(x)}p_{y}e^{-q_{i}(x)}\right)  /2\mu
\text{)},\label{9-2}\\
part3 &  =\frac{1}{\mu}\left(  e^{q_{i}(x)}p_{x}\frac{n_{i}}{R}e^{-2q_{i}%
(x)}p_{x}e^{q_{i}(x)}+e^{q_{i}(x)}p_{y}\frac{n_{i}}{R}e^{-2q_{i}(x)}%
p_{y}e^{q_{i}(x)}\right)  ,\label{9-3}%
\end{align}
with $q_{i}(x)$ satisfying the following first-order ordinary differential
equations%
\begin{align}
\frac{dq_{x}}{dx} &  =\frac{u^{\prime}}{4}\frac{2g^{2}\kappa^{3}+g\kappa
\kappa^{\prime}u^{\prime}-\kappa^{\prime\prime}}{g\kappa^{\prime}%
+\kappa^{\prime2}u^{\prime}},\label{10-1}\\
\frac{dq_{y}}{dx} &  =\frac{1}{4}\frac{2g^{2}\kappa^{3}+g\kappa\kappa^{\prime
}u^{\prime}-\kappa^{\prime\prime}}{\kappa^{\prime}-g\kappa^{2}u^{\prime}%
}.\label{10-2}%
\end{align}
These two equations (\ref{10-1})-(\ref{10-2}) have closed form solutions for
simple curves, and we list some of them below. 1) For flat plane $y=x$, we
have trivial results: $q_{x}(x)=q_{y}(x)=0$. 2)\ For a circle $y=\sqrt
{1-x^{2}}$ $(x\leq1)$, we have, $q_{x}(x)=-\left(  1/4\right)  \log(1-x^{2})$,
$q_{y}(x)=-\left(  1/2\right)  \log(|x|)$. 3)\ For a parabola $y=x^{2}$,
$q_{x}(x)=5/8\log(\left(  1+4x^{2}\right)  )$, $q_{y}(x)=\left(  5/8\right)
\log(1+4x^{2})-\left(  5/16\right)  \log(\left\vert x\right\vert )$. 4)\ For a
hyperbola $y=\sqrt{x^{2}-1}$ $(x\geq1)$, $q_{x}(x)=-1/7\log\left(
x^{2}-1\right)  +5/8\log\left(  2x^{2}-1\right)  +1/56\log\left(
6x^{2}+1\right)  ,$ $q_{y}(x)=-2/7\log\left(  x\right)  +5/8\log\left(
2x^{2}-1\right)  +1/56\log\left(  |6x^{2}-7|\right)  $. 5)\ For a sine
surface, $y=\sin x$, we have, $q_{x}(x)=5/8\log(\cos2x+3)-(1+4/\sqrt{17})/8$
$\log$ $(\left\vert 2\cos2x-\sqrt{17}-3\right\vert )-(1-4/\sqrt{17}%
)/8\log\left(  \left\vert 2\cos2x+\sqrt{17}-3\right\vert \right)  $,
$q_{y}(x)=$ $-3/10\log(\left\vert 2\cos x\right\vert )+5/8\log(3+\cos
2x)-9/40\log(7-3\cos2x)$.

One may argue that such an approach admits the geometric potential other than
$V_{g}$ (\ref{1}), the same problem as that encountered with use of the form
of constraint $df(\mathbf{x})/dt=0$. \cite{homma,ikegami} We think that this
might be a shortcoming of the use the minimum enlargement of the fundamental
commutation relations rather than use of the Dirac canonical quantization
scheme. We leave this issue for further exploration.

From either the usual centripetal force law as $a=v^{2}/R$ or the generalized
one (\ref{3}), we are tempted to conclude that the force at a point depends on
the local properties of the surface. We will see shortly, it is not the case.

\textit{Case 2: motion on a torus }Let us start from the following standard
form with\ $a\succ b\succ0$,%
\begin{equation}
f(\mathbf{x})\equiv\left(  \sqrt{x^{2}+y^{2}}-a\right)  ^{2}-b^{2}+z^{2}.
\end{equation}
This toroidal surface can be parameterized with two local coordinates
$\theta\in\lbrack0,2\pi),\varphi\in\lbrack0,2\pi),$%
\begin{equation}
\mathbf{x}=((a+b\sin\theta)\cos\varphi,(a+b\sin\theta)\sin\varphi,b\cos
\theta).\label{rr}%
\end{equation}
GCFL (\ref{2}) now gives different but equivalent forms of the right-hand side
(RHS) with consideration of the momentum $\mathbf{p}$ being perpendicular to
the normal $\nabla f(\mathbf{x})\cdot\mathbf{p=0}$,
\begin{equation}
\mathbf{\dot{p}}_{RHS}=\mathbf{-n}\frac{1}{\sin^{2}\theta}\left(  b^{3}%
K^{3}L_{z}^{2}+\frac{p_{z}^{2}}{\mu b}\right)  =\mathbf{-n}Kb\left(  \frac
{a}{b\sin^{3}\theta}\frac{p_{z}^{2}}{\mu}+\frac{\mathbf{p}^{2}}{\mu}\right)
=\mathbf{-n}Kb\left(  \frac{a}{b\sin^{3}\theta}\frac{p_{z}^{2}}{\mu}%
+2H_{c}\right)  =...\label{RHS}%
\end{equation}
where $L_{z}=xp_{y}-yp_{x}$ is the $z$-component of the angular momentum, and
$K=\sin\theta/\left(  ab+b^{2}\sin\theta\right)  $ is the Gaussian curvature
thus $\sin\theta=abK/\left(  1-b^{2}K\right)  $. Each term in any form of the
$\mathbf{\dot{p}}_{RHS}$ (\ref{RHS}) consists of two factors, $p_{i}^{2}$ (or
$L_{i}^{2}$ or $H_{c}$) and a coefficient function $Q_{i}(\mathbf{x})$ (c.f.
Eq. (\ref{last})), as it should be so from Eq. (\ref{2}). In our approach, we
take the simplest but universal way of the operator-ordering in this
term:\ $Q_{i}(\mathbf{x})p_{i}^{2}+p_{i}^{2}Q_{i}(\mathbf{x})$ when carrying
out quantization. From Eq. (\ref{RHS}). we clearly see that the generalized
centripetral force does not simply depend on the mean curvature, nor on the
Gaussian one, but interplay between curvatures and the kinematics. So far, one
might be tempted to conclude that the generalized centripetral force at a
point depends on both the kinematic quantities and the local curvatures of the
surface. In geometry for a $2D$ surface, the mean and Gaussian curvature
completely specify the local geometric properties. We will see shortly, this
is not true, either.

We take the similar way to deal with quantization, and establish the following
equations for unknown functions $q_{i}(x)$ ($i=x,y,z$),
\begin{equation}
\left[  p_{i}\mathbf{,}H\right]  =i\hbar\left(  e^{-q_{i}(x)}O\left(
\mathbf{\dot{p}}_{RHS}\right)  e^{q_{i}(x)}+e^{q_{i}(x)}O\left(
\mathbf{\dot{p}}_{RHS}\right)  e^{-q_{i}(x)}\right)  ,\text{ }(i=x,y,z),
\end{equation}
Which form of the $\mathbf{\dot{p}}_{RHS}$ is chosen, we have a set of three
differential equations up to second order for unknown functions $q_{i}(x)$
($i=x,y,z$), respectively.\ But, different forms of $\mathbf{\dot{p}}_{RHS}$
(\ref{RHS}) in quantum mechanics are not equivalent to each other. For
instance, $q_{x}(\theta,\varphi)$ with the third choice of the RHS (\ref{RHS})
$\mathbf{-n}Kb\left(  \frac{a}{b\sin^{3}\theta}\frac{p_{z}^{2}}{\mu}%
+2H_{c}\right)  $ satisfies,
\begin{align}
0 &  =8\cos\varphi(a+b\sin\theta)^{3}\left(  \cos\theta\frac{\partial
q_{x}(\theta,\varphi)}{\partial\theta}-\sin\theta(\frac{\partial q_{x}%
(\theta,\varphi)}{\partial\theta})^{2}\right)  \\
&  +8b^{3}\sin^{2}\theta\left(  \sin\varphi\frac{\partial q_{x}(\theta
,\varphi)}{\partial\varphi}-\cos\varphi(\frac{\partial q_{x}(\theta,\varphi
)}{\partial\varphi})^{2}\right)  \nonumber\\
&  -a\csc\theta\cos\varphi\left(  6a^{2}+3b^{2}+10ab\sin\theta+\left(
4a^{2}-b^{2}\right)  \cos2\theta+6ab\sin3\theta-2b^{2}\cos4\theta\right)
\nonumber
\end{align}
while $q_{x}(\theta,\varphi)$ with the first choice $\mathbf{-n}\frac{1}%
{\sin^{2}\theta}\left(  b^{3}K^{3}L_{z}^{2}+\frac{p_{z}^{2}}{\mu b}\right)  $
satisfies anther differential equation equation,
\begin{align}
0=8 &  \cos\varphi(a+b\sin\theta)^{3}\left(  (\frac{\partial q_{x}%
(\theta,\varphi)}{\partial\theta})^{2}+2\frac{\partial^{2}q_{x}(\theta
,\varphi)}{\partial\theta^{2}}\right)  \\
&  -24b^{3}\sin\theta\left(  \sin\varphi\frac{\partial q_{x}(\theta,\varphi
)}{\partial\varphi}+\cos\varphi(\frac{\partial q_{x}(\theta,\varphi)}%
{\partial\varphi})^{2}\right)  \nonumber\\
&  -\csc^{2}\theta\cos\varphi\left(  a\left(  2a^{2}+3b^{2}\right)  +3b\left(
2a^{2}-b^{2}\right)  \sin\theta-3ab^{2}\cos2\theta+b^{3}\sin3\theta\right)
\nonumber
\end{align}
The closed form solution is not available for either of the equations.
Instead, if carefully distributing the "dummy" factors, we have much a much
simpler equation with first choice $\mathbf{-n}\frac{1}{\sin^{2}\theta}\left(
b^{3}K^{3}L_{z}^{2}+\frac{p_{z}^{2}}{\mu b}\right)  $,%
\begin{equation}
\left[  p_{i}\mathbf{,}H\right]  =n_{i}\frac{b^{3}K^{3}L_{z}^{2}}{\sin
^{2}\theta}+\frac{1}{3}(pt1+pt2+pt3)
\end{equation}
where,%
\begin{align}
pt1 &  =-\frac{1}{2}\left(  e^{-\sqrt{2}q_{i}(x)}\frac{n_{i}}{\sin^{2}\theta
}\frac{p_{z}^{2}}{\mu b}e^{\sqrt{2}q_{i}(x)}+e^{\sqrt{2}q_{i}(x)}\frac
{p_{z}^{2}}{\mu b}e^{-\sqrt{2}q_{i}(x)}\frac{n_{i}}{\sin^{2}\theta}\right)
,\\
pt2 &  =-\frac{1}{2}\left(  \frac{n_{i}}{\mu b\sin^{2}\theta}p_{z}^{\prime
2}+p_{z}^{\prime2}\frac{n_{i}}{\mu b\sin^{2}\theta}\right)  ,\text{ (}%
p_{z}^{\prime2}=e^{-q_{i}(x)}p_{z}e^{2q_{i}(x)}p_{z}e^{-q_{i}(x)}\text{)},\\
pt3 &  =-e^{q_{i}(x)}p_{z}\frac{n_{i}}{\mu b\sin^{2}\theta}e^{-2q_{i}(x)}%
p_{z}e^{q_{i}(x)}.
\end{align}
Thus we find that $q_{i}(x)$ now satisfy the following first-order ordinary
differential equations, and $q_{i}(\theta,\varphi)$ are in fact independent
from $\varphi,$
\begin{align}
0 &  =16\left(  \sqrt{2}-1\right)  \cos\theta(a+b\sin\theta)^{3}\frac
{dq_{j}(\theta)}{d\theta} &  & \label{deq1}\\
+ &  \left(  2a\left(  2a^{2}+3b^{2}\right)  +6b\left(  2a^{2}-b^{2}\right)
\sin\theta-6ab^{2}\cos2\theta+2b^{3}\sin3\theta\right)  \csc\theta
,\text{(}j=x,y\text{)} &  & \nonumber\\
0 &  =4\left(  \sqrt{2}-1\right)  (\cos2\theta+3)(a+b\sin\theta)^{3}%
\frac{dq_{z}(\theta)}{d\theta} &  & \label{deq2}\\
&  +\left(  7a\left(  2a^{2}+3b^{2}\right)  +6b\left(  7a^{2}+b^{2}\right)
\sin\theta-21ab^{2}\cos2\theta-2b^{3}\sin3\theta\right)  \cot\theta. &  &
\nonumber
\end{align}
These two equations can be easily solved and the closed form solutions are
easily available but lengthy. For save the space, we only plot the the
solutions in Fig.1.

\begin{figure}
[ptb]%
\includegraphics[scale=0.9]{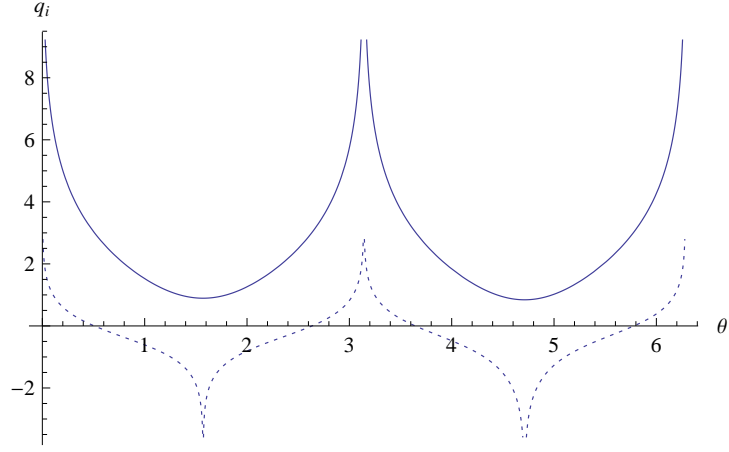}\caption{The solutions $ q_i $ for toroidal surface with $a=3,b=1$. Above (solid) curve shows $ q_x=q_y $ and below (dotted) curve shows $ q_z $}\label{figure 1}%

\end{figure}

\textit{Case 3: motion on quadric surfaces and a general 2D surface }Let us
study the following standard form with\ $a$, $b$, and $c$ taking fixed
positives, \cite{mitweb}
\begin{equation}
f(x,y,z)=\frac{x^{2}\alpha}{a^{2}}+\frac{y^{2}\beta}{b^{2}}+\frac{z^{2}\gamma
}{c^{2}}-\delta\text{, and }f(x,y,z)=0,
\end{equation}
where parameters $\alpha,\beta,\gamma$ take values either $-1$, $0$ or $1$ and
$\delta$ takes values either $0$ or $1$, depending on the classification of
quadrics containing six basic quadric surfaces such as ellipsoids,
hyperboloids of one and two sheets, elliptic cones, elliptic cylinders and
hyperbolic cylinders. \cite{mitweb} Some \textit{quadric surfaces }such as
elliptic paraboloid and hyperbolic paraboloid, etc. does not belong to the six
basic quadric surfaces, but can be treated in similar way. In this Letter, we
are only interested in cases with nonvanishing Gaussian curvature $K$, i.e.,%
\begin{equation}
K=\alpha\beta\gamma\delta\left\{  abc\left(  \frac{\left(  \alpha x\right)
^{2}}{a^{4}}+\frac{\left(  \beta y\right)  ^{2}}{b^{4}}+\frac{\left(  \gamma
z\right)  ^{2}}{c^{4}}\right)  \right\}  ^{-2}\neq0,\text{ }\alpha\beta
\gamma\delta\neq0.
\end{equation}
GCFL (\ref{2}) now gives a very compact equation,
\begin{equation}
\mathbf{\dot{p}=}-\mathbf{n}\frac{\sqrt{abc}}{\mu}\left(  \frac{K}{\alpha
\beta\gamma\delta}\right)  ^{1/4}\left(  \alpha\frac{p_{x}^{2}}{a^{2}}%
+\beta\frac{p_{y}^{2}}{b^{2}}+\gamma\frac{p_{z}^{2}}{c^{2}}\right)  .
\end{equation}
Here we like to mention another interesting appearance of the $K^{1/4}$ in
Electrostatics. The charge density of the isolated conductors whose surfaces
are quadric is also proportional to it. \cite{kmliu,chargedensity} After many
years' exploration, we know that the "local" charge density does not depend on
the local geometric properties only.\ \cite{chargedensity} The same conclusion
applies for the GCFL with a general $2D$ surface $f(\mathbf{x})=0$\textit{.}
It is straightforward to prove following equation,%

\begin{equation}
\mathbf{\dot{p}=}-\mathbf{n}\sum_{i=1}^{3}\left\{  \left(  \frac
{f_{ii}^{\prime\prime}}{f_{i}^{\prime}}+f_{i}^{\prime}\frac{f_{jk}%
^{\prime\prime}}{f_{j}^{\prime}f_{k}^{\prime}}\right)  f_{i}^{\prime}-\left(
\frac{f_{ik}^{\prime\prime}}{f_{k}^{\prime}}+\frac{f_{ij}^{\prime\prime}%
}{f_{j}^{\prime}}\right)  f_{i}^{\prime}\right\}  p_{i}^{2}\label{gen}%
\end{equation}
where the subscripts ($j,k$) in the $i$-th curly bracket $\left\{  {}\right\}
$ before $p_{i}^{2}$ differ from each other $i\neq j\neq k$. Evidently, there
is no simple relationship between the force $\mathbf{\dot{p}}$ at a point and
its curvatures of the point, with known expressions for mean and Gaussian
curvature. \cite{implicit}

During performing quantization, we find that each term of the RHS (\ref{gen})
can be divided into two noncommuting factors, $Q_{i}(\mathbf{x})\equiv
\mathbf{n}\left\{  \left(  \frac{f_{ii}^{\prime\prime}}{f_{i}^{\prime}}%
+f_{i}^{\prime}\frac{f_{jk}^{\prime\prime}}{f_{j}^{\prime}f_{k}^{\prime}%
}\right)  f_{i}^{\prime}-\left(  \frac{f_{ik}^{\prime\prime}}{f_{k}^{\prime}%
}+\frac{f_{ij}^{\prime\prime}}{f_{j}^{\prime}}\right)  f_{i}^{\prime}\right\}
$ and $p_{i}^{2}$. We find that the "dummy" factors $e^{\pm q_{i}(x)}$
allowing for both the geometric momentum and potential satisfy following three
equations,
\begin{equation}
\left[  p_{i}\mathbf{,}H\right]  =-i\hbar\sum_{i=1}^{3}\left(  e^{-q_{i}%
(x)}Q_{i}(\mathbf{x})p_{i}^{2}e^{q_{i}(x)}+e^{-q_{i}(x)}p_{i}^{2}%
Q_{i}(\mathbf{x})e^{-q_{i}(x)}\right)  ,(i=1,2,3).
\end{equation}
The explicit forms of these differential equations are available but extremely
lengthy. The closed form solutions are mathematically forbidden.

\textit{Conclusions and discussions} For the motion on the $2D$ surface, the
GCFL $\mathbf{\dot{p}=}\left[  \mathbf{p,}H_{c}\right]  $ can never be
automatically satisfied in quantum mechanics as $\left[  \mathbf{p,}H\right]
=i\hbar\left[  \mathbf{p,}H_{c}\right]  $. This difficulty originates from at
least two respects: 1) the confining technique\ implies that in quantum
mechanics, $H=\mathbf{p}^{2}/2\mu$ does no longer hold true, and has
additional curvature-induced quantum potential. 2) The interplay between the
geometric quantities and the kinematic ones $p_{i}$ and $L_{i}$ has different
forms in classical mechanics, which are not equivalent to each other in
quantum mechanics. In order to resolve the difficulty, we slightly enlarge the
canonical quantization scheme which contains not only the fundamental ones
$\left[  \mathbf{x,x}\right]  $, $\left[  \mathbf{x,p}\right]  $ and $\left[
\mathbf{p,p}\right]  $, but also $\left[  \mathbf{x,}H\right]  $ and $\left[
\mathbf{p,}H\right]  $, which constitute that minimum set to simultaneously
determine the operators $\mathbf{x,p}$ and $\mathbf{H}$. Then we demonstrate
that within the Dirac canonical quantization scheme the geometric momentum and
potential are simultaneously admissible with use of the constrain condition
$f(\mathbf{x})=0$. Thus, the difficulty is explicitly settled down for motion
constrained on the $2D$ surfaces. In addition, we find that there is no simple
relationship between the force on a point of the surface and its curvatures,
in sharp contrast to the usual centripetal force law for motion constrained on
the curve. Our work implies that within the Dirac canonical quantization
scheme, quantum mechanics for constrained motion admits more complicate forms
of curvature-induced energy. 

\begin{acknowledgments}
This work is financially supported by National Natural Science Foundation of
China under Grant No. 11175063.
\end{acknowledgments}


\begin{thebibliography}{99}
\bibitem {note}In present study, on one relation $\left[  \mathbf{p,}H\right]
=i\hbar O\left(  \left[  \mathbf{p,}H_{c}\right]  \right)  $ is explicitly utilized.
\bibitem {note2}If only the commutation relations between $[\mathbf{p,}H]$ are
required, it is not necessary to utilize the Dirac's theory of constrained systems.
\end{thebibliography}
\end{document}